\documentclass[twocolumn,pra,showpacs]{revtex4}

\begin{document}


\title{Cosmological Variation of the Fine Structure Constant versus a new Interaction}

\author{E.J. Angstmann}
\affiliation{School of Physics, University of New South Wales,
Sydney 2052, Australia}
\author{V.V. Flambaum}
\affiliation{School of Physics, University of New South Wales,
Sydney 2052, Australia}

\author{S.G. Karshenboim} 
\affiliation{D. I. Mendeleev Institute for Metrology (VNIIM), St. Petersburg 198005, Russia\\
Max-Planck-Institut f\"ur Quantenoptik, 85748 Garching, Germany
}

\date{\today}

\begin{abstract}
We show why a dynamically varying fine structure constant does not impact on the Webb group's analysis as suggested by Bekenstein. We also provide a limit on the size of a possible correction to the Dirac Hamiltonian caused by a perturbation due to a dynamically varying $\alpha$ in hydrogen.
\end{abstract}

\maketitle

\section{Introduction}
The possible variation of the fine structure constant, $\alpha$, 
is currently a very popular research topic. Webb {\textit{et al.}}
 \cite{Webb} have found evidence of $\alpha$ 
variation by analyzing absorption lines in QSO spectra. However, 
recently Bekenstein \cite{Bekenstein} has questioned the validity 
of the Webb group's analysis. Bekenstein shows that within the 
framework of dynamical $\alpha$ variability the form of the Dirac 
Hamiltonian relevant for an electron in an atom departs from the 
standard form. Unfortunately no self consistent quantum 
electrodynamic theory was derived. Instead the Dirac 
Hamiltonian, \^H, was presented for an electron bound by a 
Coulomb field:  
\begin{eqnarray}\label{eq:H}
\hat{H} & = & \hat{H}_{0}+\delta\hat{H} \\
\label{eq:H_0}\hat{H}_{0} & = & (-\imath\hbar c\mbox{\boldmath$\alpha$}\cdot\mbox{\boldmath$\nabla$}+mc^2\beta+e\Phi I)  \\
\label{eq:deltaH}\delta\hat{H} & = & (I-\beta)\tan^2\chi\cdot V_{C}
\end{eqnarray}
where $V_{C}=-Z\alpha/r$. The last term is related to an
 effective correction to the Coulomb field due to the dynamical 
nature of $\alpha$, and $\tan^{2}\chi$ is a small parameter. In 
other words, a dynamically varying fine structure constant can be 
accounted for as a perturbation of the Dirac Hamiltonian. This 
perturbative term, $\delta\textrm{\^H}$, vanishes in a non-relativistic 
approximation but it produces some relativistic corrections which 
can be studied both in astrophysical spectra and laboratory conditions.

In the following sections we will show how this perturbation shifts 
atomic energy levels. In particular we pay attention to heavy atoms 
which provide us with astrophysical data and are the most sensitive to a 
possible $\alpha$ variation. We also consider atomic hydrogen, 
since it is the best understood atomic system for laboratory experiments.

\section{Heavy Atoms}
Heavy atoms are of interest to us since they are the most sensitive to a 
varying $\alpha$. We performed a calculation to show how the modified
 form of the Dirac Hamiltonian effects the energy of an external electron 
in a heavy atom. We averaged $\delta$\^H, presented in Eq. 
(\ref{eq:deltaH}), over the relativistic wave 
function for electrons near the nucleus \cite{Khriplovich} at 
zero energy. The use of the wave function for the electrons 
at zero energy is justified by noting that the main contribution 
to the electron's energy comes from distances close to the 
nucleus, $r \sim a/Z$ \cite{Landau}. At a distance of one Bohr radius, 
$r \sim a$, the potential is screened by the other electrons 
and the potential energy is given by $V_{C} \sim m\alpha^{2}$. 
This is of the same order of magnitude as the binding energy 
of the electron, $E \sim m\alpha^{2}$. However, inside the 
Bohr radius at $r \sim a/Z$, screening is negligible and $
V_{C}\sim Z^{2}m\alpha^{2}$. Inside this region $V_{C}\gg E$ 
and so the binding energy of the electron can be safely ignored.

The non-relativistic limit of $\delta E$ was taken. This gave:
\begin{equation}\label{eq:heavyatomcorrect}
\frac{\delta E}{E} = \frac{2(Z\alpha)^{2}\tan^{2}\chi}{\nu}\frac{1}{j + 1/2}
\end{equation}
where $E=-Z_{a}^{2}mc^{2}\alpha^{2}/(2\nu^{2})$ is the energy of the electron, and $Z_{a}$ is the 
charge ``seen'' by the electron - it is 1 for atoms, 2 for singly 
charged ions etc.

It is interesting to note that this correction to the energy of
 the electron has exactly the same form as the relativistic 
correction, $\Delta$, to the energy of an external electron:
\begin{equation}\label{eq:relcorrect}
\frac{\Delta}{E} = \frac{(Z\alpha)^{2}}{\nu}\frac{1}{j + 1/2}.
\end{equation}
This makes the effect of the modified form of the Dirac 
Hamiltonian indistinguishable from a small change in
 $\alpha^{2}$ in Eq. (\ref{eq:relcorrect}). 
Since Eq. (\ref{eq:heavyatomcorrect}) and (\ref{eq:relcorrect}) 
are directly proportional, and $\tan^{2}\chi$ is necessarily 
small, there is no need for any modification to the Webb group's 
analysis in heavy atoms since $\tan^{2}\chi$ is 
accommodated into the change of $\alpha$.

In this derivation we assumed 
that we could consider the unscreened Coulomb field, this is 
clearly not the case for a valence electron in a many electron 
atom. We justify this assumption by once again noting that the main 
contribution to the energy is given by distances close to the 
nucleus, of the order of the Bohr radius over $Z$, $r \sim a/Z$.
 At this distance the 
only screening comes from the $1s$ electrons and we can use 
Slater's rules to give the more correct version of Eq. 
(\ref{eq:heavyatomcorrect}) and (\ref{eq:relcorrect}):
\begin{eqnarray}
\frac{\delta E}{E} & = &  \frac{2\alpha^{2}(Z-0.6)^{2}\tan^{2}\chi}{\nu}\frac{1}{j + 1/2} \\
\frac{\Delta}{E} & = & \frac{\alpha^{2}(Z-0.6)^{2}}{\nu}\frac{1}{j + 1/2}.
\end{eqnarray}
This describes non hydrogen-like or helium-like ions or atoms. This does not 
effect the proportionality of the two terms and makes 
very little difference to the energies in heavy atoms. 
Consideration of many-body corrections has shown that 
this does not change the proportionality relationship either. 

\section{Calculations Involving Hydrogen Atom}

The case is somewhat simplified for the hydrogen atom and other
 hydrogen-like ions as there are no inter-electron interactions. 
There are also very accurate experimental measurements of 
transition frequencies in hydrogen.

We confirm Bekenstein's result \cite{Bekenstein} that applying 
the Hamiltonian (\ref{eq:H}) one can derive:
\begin{equation} \label{eq:deltaE}
\delta E = -\frac{mc^{2}Z^{4}\alpha^{4}}{n^{3}}\left(\frac{1}{j+1/2}-\frac{1}{2n}\right) \tan^{2}\chi \;.
\end{equation}
Note that for large $n$ (zero energy), this shift is again proportional to the relativistic correction, $\Delta$, given by:
\begin{equation}
\Delta = -\frac{mc^{2}Z^{4}\alpha^{4}}{2n^{3}}\left(\frac{1}{j+1/2}-\frac{3}{4n}\right).
\end{equation}

The most accurate measurements in hydrogen atoms are related to $1s-2s$, $2s-ns$ and $2s-nd$ transitions with $n=8, 10, 12$. However, the data cannot be used directly to find $\delta E$, the difference between the experimental and theoretical energy of a level, since it is necessary to determine a value of the Rydberg constant and the $1s$ Lamb shift. The latter can be calculated theoretically with a relatively large uncertainty due to the proton radius \cite{radius} and the uncertainty of a direct experimental determination is also quite high. If one intends to extract a value of $\tan^{2}\chi$, data of at least three accurate measurements should be used to obtain a self consistent answer. However, in addition to the $1s-2s$ transition, only data of the $2s-ns/d$ transitions are available with a comparable accuracy (see the compilation \cite{Mohr}), and they are only very weakly sensitive to the value of $n$, due to the $1/n^{4}$ scaling, and thus the sensitivity of such a test is quite low.

Fortunately there is another approach which needs neither a value of the Rydberg constant nor the $1s$ Lamb shift. It is based on a comparison of theoretical and experimental data for the $2p_{3/2}-2p_{1/2}$ splitting. The experimental value
\begin{equation}
f_{2p_{3/2}\rightarrow2p_{1/2}}({\rm exp})=10\,969\,045(15)\, \textrm{kHz}
\end{equation}
is derived from two experimental results,
\begin{eqnarray}
f_{2s_{1/2}\rightarrow2p_{3/2}}({\rm exp}) & = & 911\,200(12)\,\textrm{kHz}\\
f_{2s_{1/2}\rightarrow2p_{1/2}}({\rm exp}) & = & 1\,057\,845(9)\, \textrm{kHz}
\end{eqnarray}
presented in \cite{Lundeen} and \cite{Hagley} respectively. It has to be compared with a theoretical value which we take from a compilation \cite{Mohr} (see also review \cite{Eides})
\begin{equation}
f_{2p_{3/2}\rightarrow2p_{1/2}}({\rm theory})=10\,969\,041.2(1.5)\, \textrm{kHz}\;.
\end{equation}
Now using Eq. (\ref{eq:deltaE}) we can write:
\begin{equation}
f_{2p_{3/2}\rightarrow2p_{1/2}}({\rm exp})-f_{2p_{3/2}\rightarrow2p_{1/2}}(\rm theory)=\frac{mc^{2}\alpha^4}{16h}\tan^{2}\chi.
\end{equation}
Noting that the leading contribution to $f_{2p_{3/2}\rightarrow2p_{1/2}}(\rm theory)$ is given by $mc^{2}\alpha^4/32h$ allows us to write:
\begin{equation}
f_{2p_{3/2}\rightarrow2p_{1/2}}({\rm exp})=f_{2p_{3/2}\rightarrow2p_{1/2}}({\rm theory})(1+2\tan^{2}\chi)
\end{equation}
Using the values above, this tells us that $\tan^{2}\chi=2(7)\times10^{-7}$. This is consistent with an $\alpha$ that is not varying dynamically within this framework.

\section{Conclusions}

In conclusion, using the modified form of the Dirac Hamiltonian does not 
effect the Webb group's analysis. The Webb group's results can not 
distinguish between the $\alpha$ variability from phase transitions 
and dynamical $\alpha$ variability. The most stringent limit we can 
place on Bekenstein's \cite{Bekenstein} $\tan^{2}\chi$ parameter 
is $\tan^{2}\chi=2(7)\times 10^{-7}$. This is consistent with no 
dynamical $\alpha$ variability according to the framework laid 
out in Bekenstein's paper \cite{Bekenstein}. 

We should also mention that such a weak limitation was obtained only because we had to apply an effective operator $\delta\hat{H}$ which vanishes in a leading non-relativistic approximation. A self consistent quantum electrodynamic theory with a dynamically varying $\alpha$ should meet some even stronger constraints due to a comparison of the value of the fine structure constant from the anomalous magnetic moment of the electron ($\alpha^{-1}=137.035\,998\,80(52)$ \cite{kinoshita}) which should have a correction of fractional order of $\tan^{2}\chi$ and other values of $\alpha$ which are mostly derived via a complicated chain of relations with $\alpha$ eventually coming from the Rydberg constant and thus quite weakly affected by $\delta\hat{H}$ (the fractional value of the correction is to be of order of $\alpha^2\tan^{2}\chi$) and hence allowing one to extract $\delta \alpha/\alpha \sim \tan^{2}\chi$.  Such a comparison will likely lead to a limitation on $\tan^{2}\chi$ at a level of a few parts in $10^{-8}$. E.g. the most accurate result obtained this way is $\alpha^{-1}=137.036\,000\,3(10)$ \cite{chu} and thus  
$\delta \alpha/\alpha = 11(8)\times 10^{-9}$. Another set of questions due to a modified version of QED should target its gauge invariance, renormalizability and Ward identities, which supports the same charge for electrons and protons. The current QED construction is quite fragile and it is not absolutely clear if it can be successfully extended. All these questions need to be answered clearly before any modification of QED is considered seriously.

\section*{Acknowledgments}
The work of SGK was supported in part by RFBR grant 03-02-16843. A part of the work was done during a short but fruitful visit of VVF to MPQ. The work of EJA and VVF was supported by the Australian Research Council.

\end{document}